\documentclass[options]{kluwer}

\usepackage{graphics}
\usepackage{epsf,epsfig}

\newcommand{\km}{\,{\rm km}}

\begin{document}
\begin{article}

\begin{opening}
\title{Numerical simulations of the accretion-ejection instability}

\author{S.~E. Caunt}
\author{M.\ Tagger}
\institute{DSM/DAPNIA/Service d'Astrophysique (CNRS URA 2052), C.E.A.
Saclay, 91191 Gif sur Yvette, France}

\runningauthor{Caunt \& Tagger}
\runningtitle{Numerical simulations of the AEI}

\begin{abstract} 

The Accretion-Ejection Instability (AEI) is explored numerically using a
global 2d model of the inner region of a magnetised accretion disk.  The
disk is initially currentless but threaded by an external vertical
magnetic field created by external currents, and frozen in the flow.  In
agreement with the theory a spiral instability, similar in many ways to
those observed in self-gravitating disks, but driven by magnetic
stresses, develops when the magnetic field is close to equipartition
with the disk thermal pressure.  The present non-linear simulations give
good evidence that such an instability can occur in the inner region of
accretion disks.


\end{abstract}

\end{opening}

%---------------------------------------------------------------

\section{Introduction}

% As shown by , 
The accretion-ejection instability (AEI) (Tagger \& Pellat
\cite{TaggerP99}; see also Varni\`ere et al., Rodriguez et al., and
Tagger et al., this workshop) can occur close to the inner edge of a
magnetised accretion disk when the plasma beta (the ratio of gas to
magnetic pressures) is around unity.
% for which the quantity $\Omega\Sigma/B_0^2$
% increases radially where $\Omega$ is the angular velocity, $\Sigma$ the
% surface density and $B_0$ the magnetic field for which 
The instability appears, if the magnetic field decreases outwards
sufficiently fast, as a spiral wave of low azimuthal wavenumber, made
unstable through the interaction with a Rossby vortex (associated with
the gradient of vorticity) which it generates at its corotation radius.
% This also
% causes the emission of Alfv\'en waves from corotation vertically along
% magnetic field lines, as their footpoints are twisted .

We have undertaken numerical simulations, with a setup optimised
according to our previous knowledge of the AEI. We consider an
infinitely thin disk threaded by a moderate vertical magnetic field:
this is justified by the properties of the AEI, which are essentially
constant vertically across the disk (unlike the magneto-rotational
instability (Balbus \& Hawley, 1991) for which vertical modes are
required).  The disk is embedded in vacuum and, in order to separate
different physical effects, we consider here only configurations where
initially the equilibrium magnetic field is due to external currents. 
Starting with small scale initial random fluctuations in the fluid
velocity, the simulations show large-scale spiral perturbations evolving
in the disk, and corresponding to the expected properties of the AEI.

The following sections present the essentials of the model, basic
results and conclusions. More detailed discussions of these are covered
elsewhere (Caunt \& Tagger, 2000).

%---------------------------------------------------------------

%%%%%%%%%%%%%%%%%%%%%%%%%%%%%%%%%%%%%%%%%%%%%%%%%%%%%%%%%%%%%%%%%%%%%

%---------------------------------------------------------------
\section{The model}
%---------------------------------------------------------------

Under the assumption, valid for the physics of the AEI, that
perturbations are essentially independent of $z$ in the disk, we
integrate vertically the standard MHD equations (momentum, continuity,
induction) describing the fluid motion in a disk around a central
object.  We thus solve for the 2d ($r,\phi$) evolution of the fluid
properties (surface density, velocity, magnetic field) averaged over
$z$.  A conservative scheme of the type described by Stone \& Norman
\cite{StoneN92} is used to ensure that mass, angular momentum and
magnetic flux are conserved to numerical precision throughout the
simulation.

We assume that the disk is threaded by a poloidal magnetic field which
is symmetric about the midplane, hence purely vertical at $z=0$.  In the
vacuum surrounding the disk the magnetic field can be described by a
magnetic potential which, in turn, results from perturbed currents in
the disk.  It can thus be calculated from a Poisson equation, similar to
the one describing the gravitational potential of a self-gravitating
disk, but whose source is the vertical component of the magnetic field
at the disk surface (rather than density in the gravitational case). 
Currents within the disk can be derived from the jump in the horizontal
component of the field across the disk.  Hence magnetic stresses can be
included in an otherwise purely hydrodynamic model.

We assume that the central object has a mass of $M=10M_\odot$ and the
disk extends from between $r=1,000\km$ to $r=50,000\km$ with a constant
aspect ratio of $\epsilon=0.1$.  We use a logarithmic radial grid which
allows us to get a better resolution in the inner region of the grid,
where it is necessary, and to model a very large radial extent of the
disk.  It also has the added advantage that unwanted effects of the
boundary condition at the outer radius of the simulation are avoided
(waves are damped more effectively as they approach the outer radius).

%---------------------------------------------------------------

%%%%%%%%%%%%%%%%%%%%%%%%%%%%%%%%%%%%%%%%%%%%%%%%%%%%%%%%%%%%%%%%%%%%%

%---------------------------------------------------------------
\section{Results}
%---------------------------------------------------------------

\begin{figure}
\begin{center}
\epsfig{file=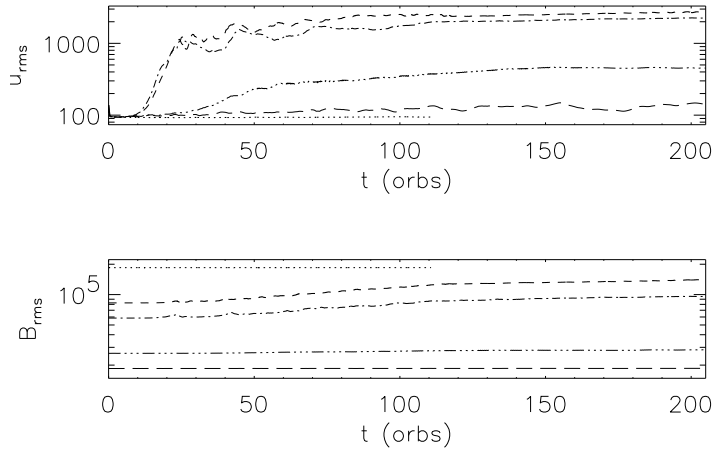,width=0.8\hsize} 
\caption{Comparison of the
evolution of the instability for different values of the initial $\beta$.
Values of $\beta=0.1$, $0.5$, $1.0$, $5.0$ and $10.0$ are shown as
dotted, dot-dashed, dashed, tripple-dot-dashed and long-dashed lines
respectively.  Time is given in units of the orbital time at the inner
radius. The plots show the evolution of (top) the rms velocity, and
(bottom) the rms magnetic field.}
\label{f:urmscomp}
\end{center}
\end{figure}

The results presented here are aimed at illustrating certain
properties of the instability as determined by linear theory,
specifically the growth rate/magnitude of instability for different
field strengths and development of the spiral wave and Rossby vortex.

Figure \ref{f:urmscomp} shows the evolution of the rms velocity within
the disk for values of $\beta=0.1$, $0.5$, $1.0$, $5.0$ and $10.0$.  As
is shown, the perturbed velocity grows rapidly for $\beta=0.5$ and
$\beta=1.0$, exceeding the initial random noise after approximately 10
orbits.  The growth is approximately exponential during this time.
For the other values of $\beta$ we see very little action in
comparison.  The $\beta=0.1$ and $\beta=10.0$ cases are virtually
stable with the $\beta=5.0$ being somewhat more active but failing to
achieve the same amplitude as the most unstable cases.  This run is
still interesting to observe as the growth of the instability is
slower and the evolution of the spiral wave is somewhat cleaner.

%-----------------------------------------------------
\begin{figure}
\newlength\ltest
\ltest=0.98\hsize
\divide\ltest by 3
\epsfig{file=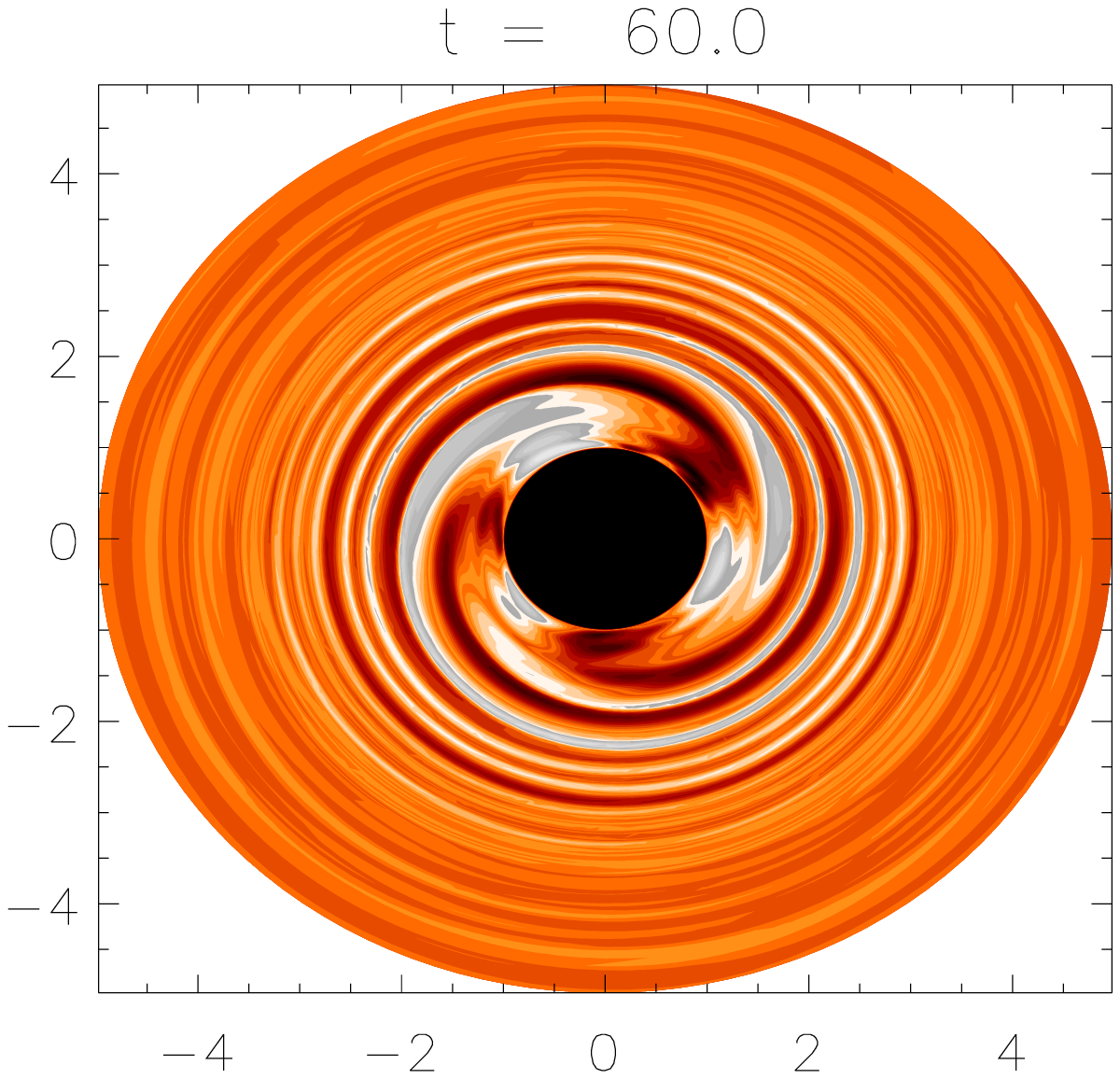,width=\ltest}  
\epsfig{file=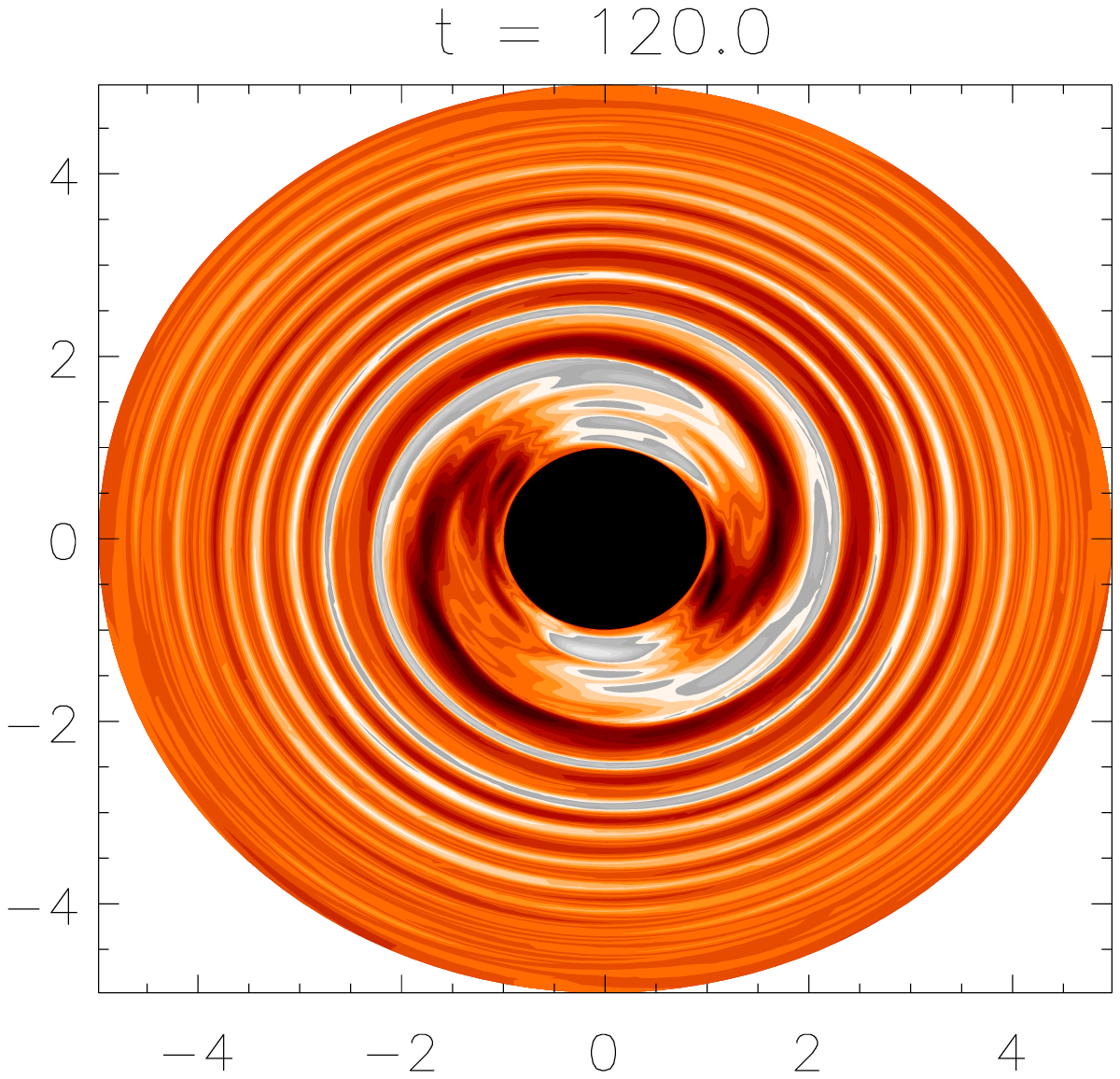,width=\ltest}  
\epsfig{file=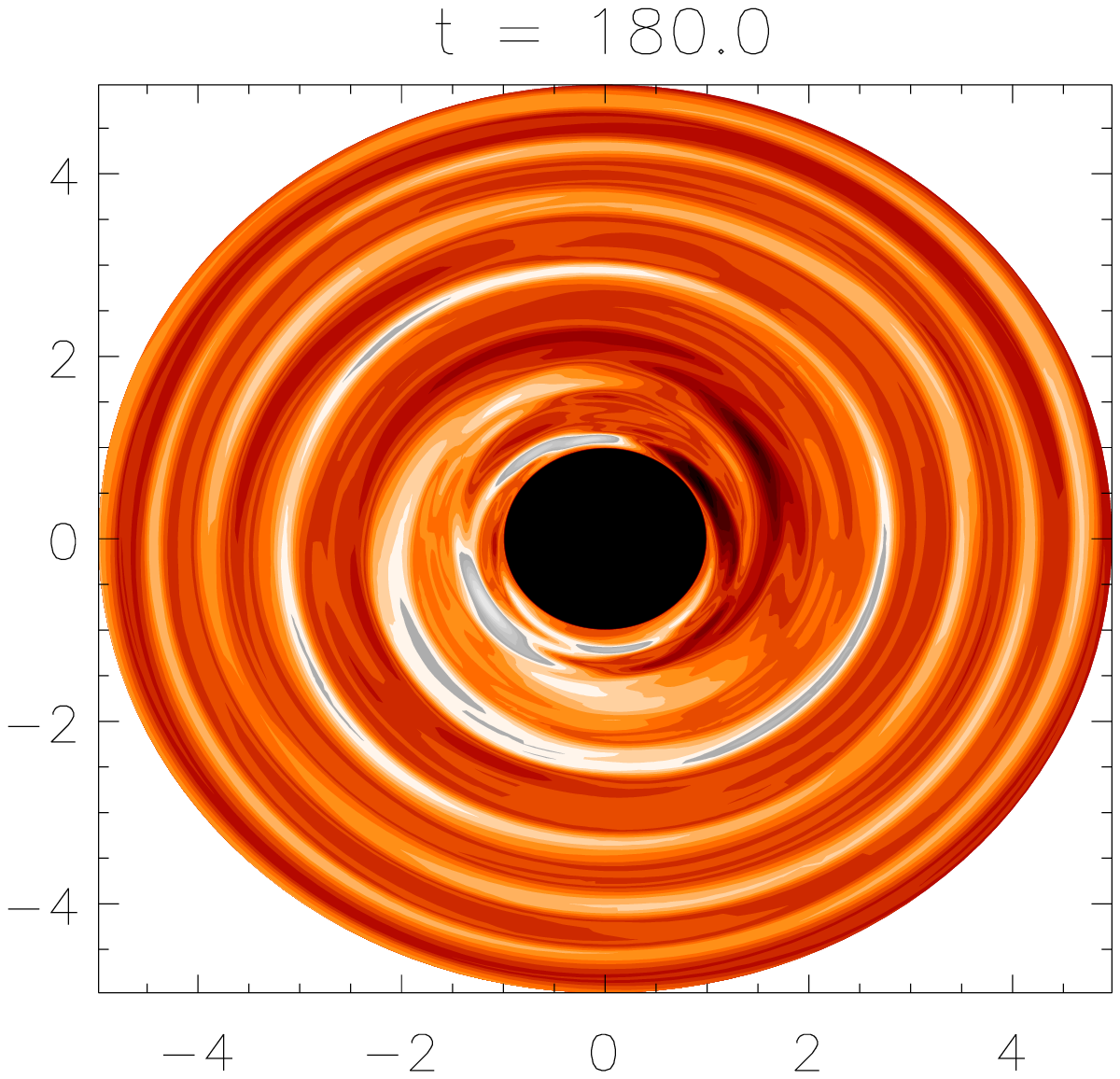,width=\ltest}
\caption{Plots of the spiral structure in the inner region of the disk
at times $t=60$, $120$ and $180$ orbits. Starting from initially
random fluctuations at $t=0$, the disk has clearly developed 3-armed
followed by 2-armed and finally 1-armed spirals.}
\label{f:b=5.sprl}
\end{figure}
%-----------------------------------------------------

The evolution of the instability into a well developed spiral wave is
best seen for the $\beta=5.0$ case. Figure \ref{f:b=5.sprl} shows the
different number of spiral arms that occur over time (plotted is the
radial velocity): initial random perturbations develop to $m=3$, $m=2$
and $m=1$ spiral structures, as the magnetic flux is advected toward the 
central region. The transition is generally through a
mixture of these for all values of unstable $\beta$. For the more
active values ($\beta=0.5$ and $\beta=1.0$) the transition is much
quicker and always ends up with $m=1$.

%-----------------------------------------------------
\begin{figure}
\ltest=0.98\hsize
\divide\ltest by 2
\epsfig{file=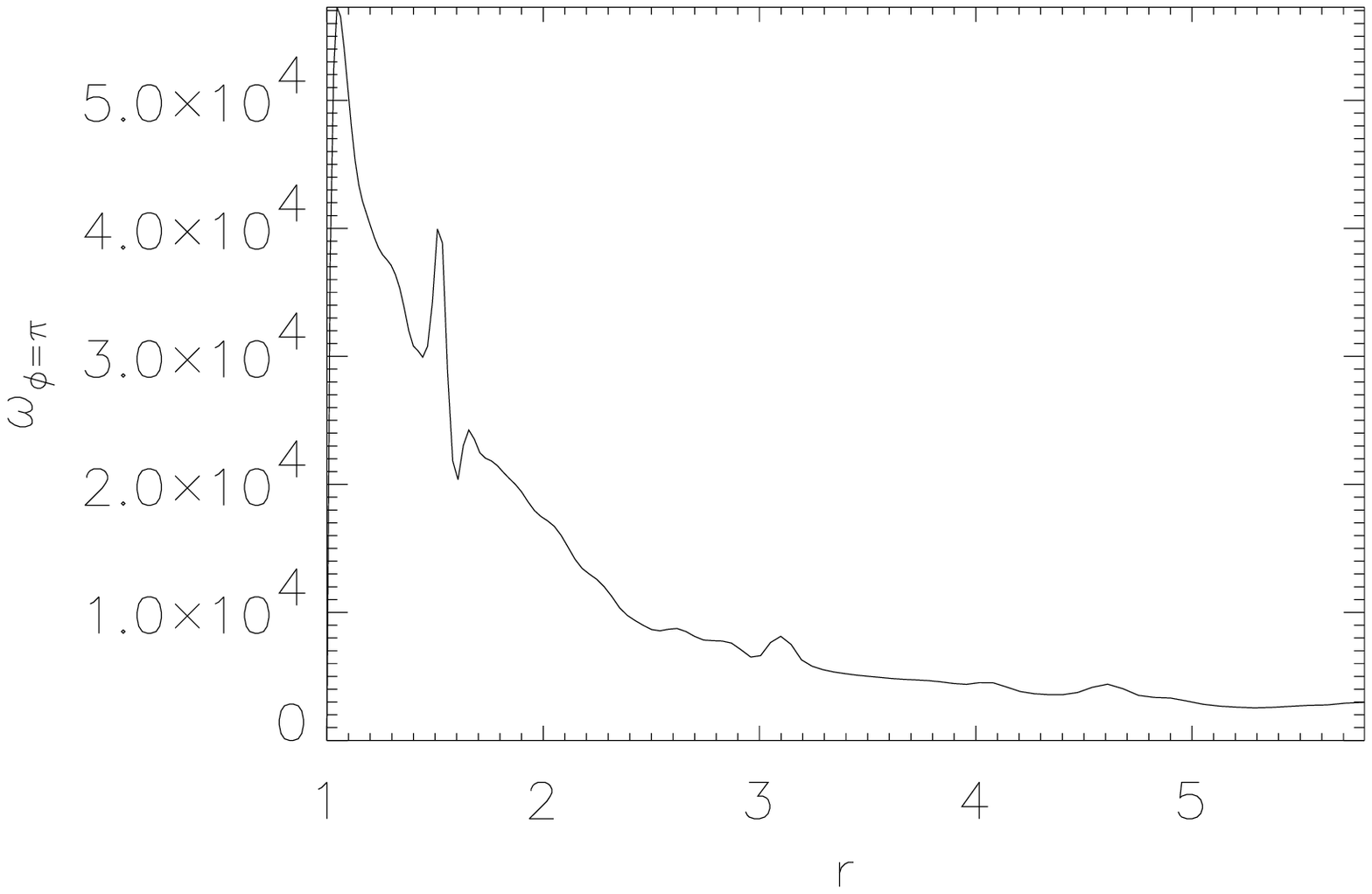,width=\ltest}
\epsfig{file=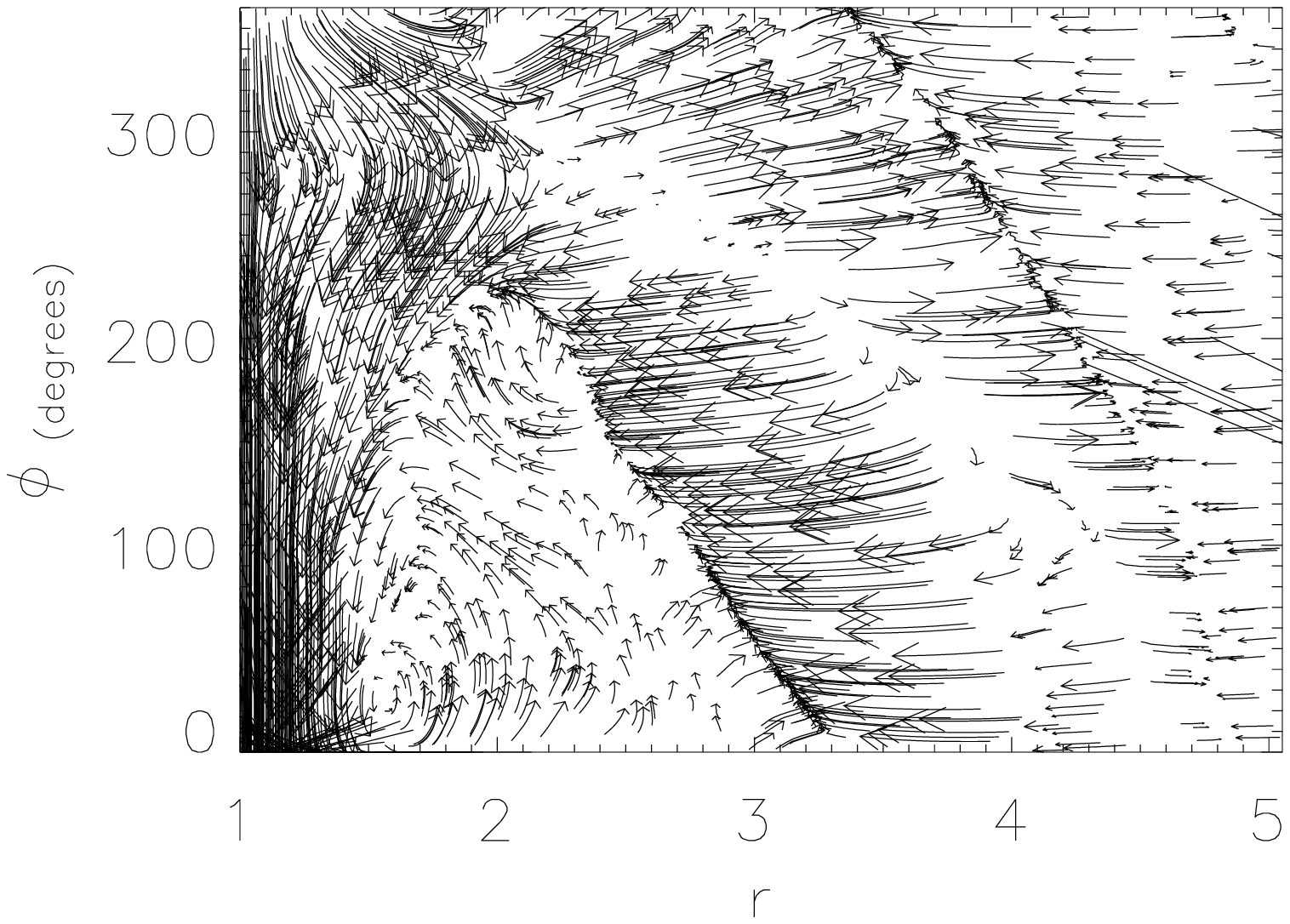,width=\ltest} 
\caption{(left hand): Vorticity in the inner region of the
accretion disk. (right hand): The velocity
field within the disk in Cartesian projection. The vortex is clearly
visible around $r\sim 1.5$.}
\label{f:b=0.5.vortic2}
\end{figure}
%-----------------------------------------------------

The AEI relies on the existence of a Rossby vortex generated at
co-rotation. Figure \ref{f:b=0.5.vortic2}a shows the radial variation
of vorticity in the disk. Clearly shown is a discontinuity close to
the inner radius. The velocity field, shown in Figure
\ref{f:b=0.5.vortic2}b clearly shows that the fluid flow contains a
vortex at the same radius. This is a good indication that a Rossby
vortex has indeed been generated at corotation.

%%%%%%%%%%%%%%%%%%%%%%%%%%%%%%%%%%%%%%%%%%%%%%%%%%%%%%%%%%%%%%%%%%%%%

%---------------------------------------------------------------

\section{Conclusions}

Using this model we have shown the essential features predicted by
linear theory of the AEI. We have seen that the instability is clearly
strongest for field strengths of the order of equipartition, and leads
to the development of low azimuthal wavenumber spiral waves and a Rossby
vortex.

Future work will link in the third type of wave (namely the vertical
emission of Alfv\'en waves) from which the formation of winds and jets
is believed to be possible.

%---------------------------------------------------------------

%---------------------------------------------------------------

\end{article}
\end{document}